\documentclass[twocolumn,prl,showpacs,preprintnumbers,amsmath,amssymb]{revtex4}
\usepackage{graphicx}
\usepackage{mathptmx,bm}

\begin{document}


\title{Photoemission and X-ray Absorption Studies of the Diluted Magnetic Semiconductor Ba$_{1-x}$K$_{x}$(Zn$_{1-y}$Mn$_{y}$)$_{2}$As$_{2}$ Isostructural to Fe-based Superconductors}

\author{H. Suzuki$^{1}$, K. Zhao$^{2}$, G. Shibata$^{1}$, Y. Takahashi$^1$, S. Sakamoto$^{1}$, K. Yoshimatsu$^{1}$, B. J. Chen$^{2}$, H. Kumigashira$^{3}$, F.-H. Chang$^4$, H.-J. Lin$^4$, D. J. Huang$^4$, C. T. Chen$^4$, Bo Gu$^5$, S. Maekawa$^5$, Y. J. Uemura$^{6}$, C. Q. Jin$^{2}$ and  A. Fujimori$^{1}$}
 
\affiliation{$^{1}$Department of Physics, University of Tokyo,
Bunkyo-ku, Tokyo 113-0033, Japan}

\affiliation{$^{2}$Beijing National Laboratory for Condensed Matter Physics, Institute of Physics, Chinese Academy of Sciences, Beijing 100190, China}

\affiliation{$^{3}$KEK, Photon Factory, Tsukuba, Ibaraki 305-0801, Japan}

\affiliation{$^{4}$National Synchrotron Radiation Research Center, Hsinchu 30076, Taiwan}

\affiliation{$^{5}$Advanced Science Research Center, Japan Atomic Energy Agency, Tokai 319-1195, Japan}

\affiliation{$^{6}$Department of Physics, Columbia University, New York, New York 10027, USA}
\date{\today}

\begin{abstract}
The electronic and magnetic properties of a new diluted magnetic semiconductor (DMS) Ba$_{1-x}$K$_{x}$(Zn$_{1-y}$Mn$_{y}$)$_{2}$As$_{2}$, which is isostructural to so-called 122-type Fe-based superconductors, are investigated by x-ray absorption spectroscopy (XAS) and resonance photoemission spectroscopy (RPES). Mn $L_{2,3}$-edge XAS indicates that the doped Mn atoms have the valence 2+ and strongly hybridize with the $4p$ orbitals of the tetrahedrally coordinating As ligands. The Mn $3d$ partial density of states (PDOS) obtained by RPES shows a peak around 4 eV and relatively high between 0-2 eV below the Fermi level ($E_{F}$) with little contribution at $E_{F}$, similar to that of the archetypal DMS Ga$_{1-x}$Mn$_{x}$As. This energy level creates $d^{5}$ electron configuration with $S=5/2$ local magnetic moments at the Mn atoms. Hole carriers induced by K substitution for Ba atoms go into the top of the As $4p$ valence band and are weakly bound to the Mn local spins. The ferromagnetic correlation between the local spins mediated by the hole carriers induces ferromagnetism in Ba$_{1-x}$K$_{x}$(Zn$_{1-y}$Mn$_{y}$)$_{2}$As$_{2}$.

\end{abstract}

\pacs{75.50.Pp,74.70.Xa,79.60.-i,78.70.Dm}

\maketitle

Diluted magnetic semiconductors (DMSs) have received much attention due to the possibility of utilizing both charge and spin degrees of freedom in electronic devices \cite{Ohno.H_etal.Science1998,Dietl.T_etal.Nat-Mater2010,Zutic.I_etal.Rev_mod_Phys2004,Ohno.H_etal.Applied-Physics-Letters1996,Dietl.T_etal.Science2000}. In order to realize functional spintronics devices, it is important to have a full control of the carrier density and ferromagnetic Curie temperature ($T_{C}$). Prototypical DMS systems such as Ga$_{1-x}$Mn$_{x}$As, In$_{1-x}$Mn$_{x}$As and Ga$_{1-x}$Mn$_{x}$N, however, show severely limited chemical solubility due to the substitution of divalent Mn atoms for the trivalent Ga or In sites. Besides, the simultaneous doping of charge and spin induced by Mn substitution prevents us from optimizing the charge and spin densities independently, although the tuning of the charge density by applying gate voltage is possible and can be used to control $T_{C}$ \cite{Sawicki.M_etal.Nat-Phys2010}. 

A newly-found DMS, Ba$_{1-x}$K$_{x}$(Zn$_{1-y}$Mn$_{y}$)$_{2}$As$_{2}$ \cite{Zhao.K_etal.Nat-Commun2013} (Mn-BaZn$_{2}$As$_{2}$), is isostructural to the ``122''-type iron-based high-temperature superconductors \cite{Paglione.J_etal.Nat-Phys2010} and has a $T_{C}$ as high as $230$ K \cite{Zhao.K_etal.Chin.-Sci.-Bull.2014}. This material has an advantage that the charge reservoir Ba layer and the ferromagnetic ZnAs layer are spatially separated, which allows us to control the amount of hole carriers by K substitution to the Ba layer and that of magnetic elements by substituting Mn to the ZnAs layer rather independently. In addition, the substitution of Mn atoms for isovalent Zn atoms enables us to circumvent the difficulty of limited chemical solubility in Ga$_{1-x}$Mn$_{x}$As and related DMSs, which makes it possible to obtain bulk specimens. Anomalous Hall effect observed in Mn-BaZn$_{2}$As$_{2}$ \cite{Zhao.K_etal.Nat-Commun2013} provides evidence that ferromagnetism here is intrinsic as in Ga$_{1-x}$Mn$_{x}$As \cite{Ohno.H_etal.Phys.-Rev.-Lett.1992,Jungwirth.T_etal.Phys.-Rev.-Lett.2002}. This new series of DMSs, together with the new ``111''-type materials Li(Zn,Mn)As  \cite{Deng.Z_etal.Nat-Commun2011} and Li(Zn,Mn)P \cite{Deng.Z_etal.Phys.-Rev.-B2013}, opens up new possibilities for the next generation spintronics devices. To achieve this goal, it is important to investigate the electronic and magnetic structure of Mn-BaZn$_{2}$As$_{2}$ and clarify the advantages of this material as compared with the archetypal Mn-doped-based DMS materials like Ga$_{1-x}$Mn$_{x}$As and In$_{1-x}$Mn$_{x}$As.

In the present work, we have performed x-ray absorption spectroscopy
(XAS) and resonance photoemission
spectroscopy (RPES) measurements on Mn-BaZn$_{2}$As$_{2}$ ($x=0.3$, $y=0.15$, $T_{C}=180$ and 230 K). Polycrystalline samples were synthesized under high pressure by the method described in Ref. \onlinecite{Zhao.K_etal.Nat-Commun2013}. XAS measurements for samples with $T_{C}=230$ K were performed at the Dragon Beamline BL-11A of National Synchrotron Radiation Research Center (NSRRC), Taiwan. The spectra were taken in the total-electron yield (TEY: probing depth $\sim$ 5 nm) mode. The monochromator resolution was $E/\Delta E>10000$ and the x-rays were circularly polarized. The samples were filed \textit{in situ} before the measurements to obtain fresh surfaces. RPES experiments for samples with $T_{C}=180$ K were performed at Beamline 2C of Photon Factory, High-Energy Accelerator Research Organization. RPES experiments were done before the synthesis of $T_{C}=230$ K samples \cite{Zhao.K_etal.Chin.-Sci.-Bull.2014}. Calibration of the Fermi level ($E_{F}$) was achieved using the $E_{F}$ of gold which was in electrical contact with the samples. Incident photon energies from 635 eV to 643 eV were linearly polarized. 
In order to gain insight further into the electronic structure of the host semiconductor BaZn$_{2}$As$_{2}$, we have performed density-functional theory (DFT) calculations. 

\begin{figure}[htbp] 
   \centering
   \includegraphics[width=8cm]{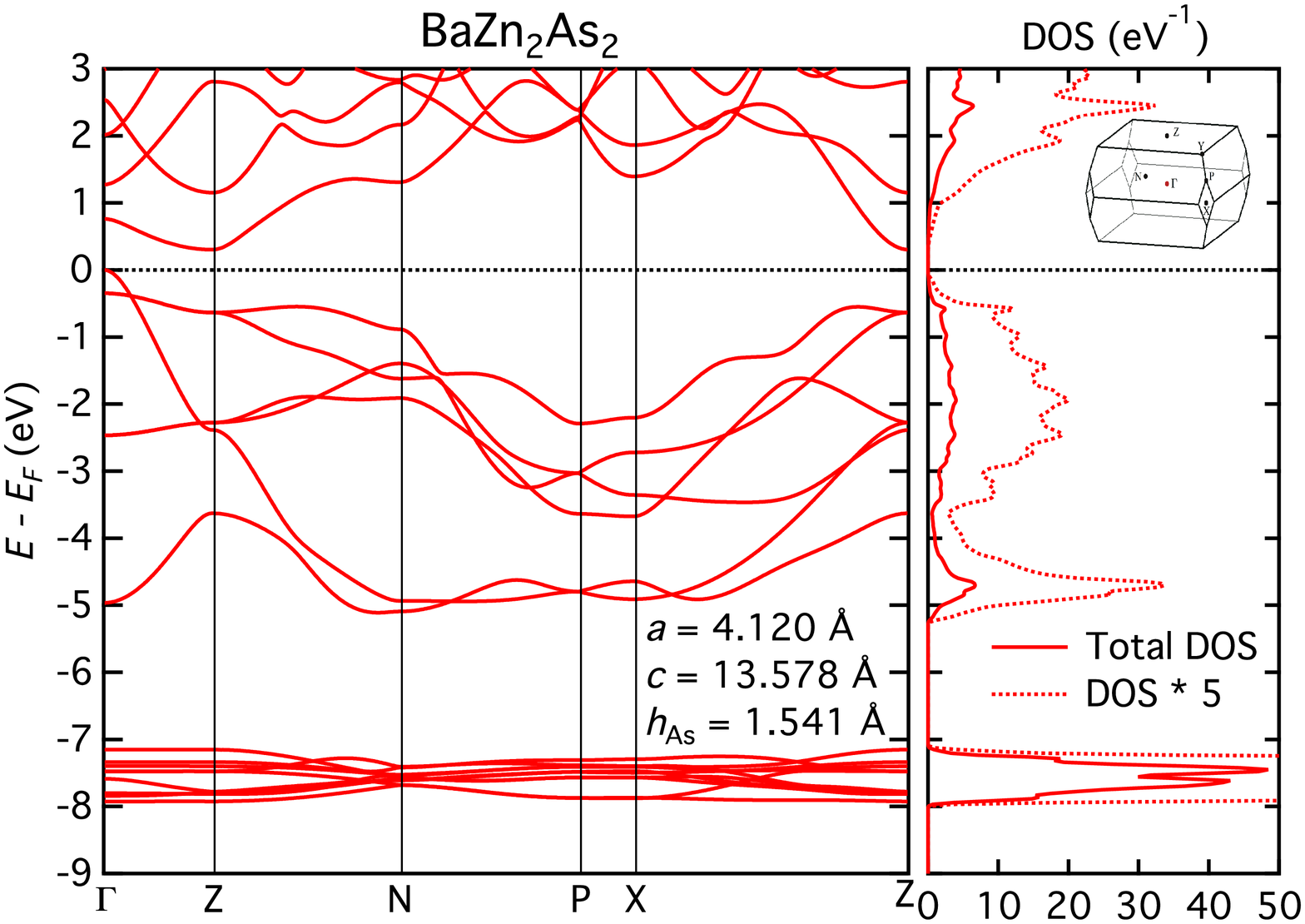} 
   \caption{(Color online) Band structures and the density of states for BaZn$_{2}$As$_{2}$ calculated using modified Becke-Johnson exchange potential. The lattice parameters $a=4.120$ \AA\ , $c=13.578$ \AA\ \cite{Zhao.K_etal.Nat-Commun2013} and $h_{\text{As}}=1.541$ \AA\ \cite{Hellmann.A_etal.Z.-Naturforsch.2007} are used in the calculation.} 
   \label{LDAband}
\end{figure}

To understand the electronic structure of the host semiconductor, we performed band-structure calculations on BaZn$_{2}$As$_{2}$ using the Wien2k package \cite{Blaha.P_etal.2001}. Figure \ref{LDAband} (a) shows the calculated band structure and the density of states (DOS). BaZn$_{2}$As$_{2}$ has space group I4/mmm and the first Brillouin zone is shown in the inset of the right panel. The calculations were done using the experimentally determined tetragonal lattice constants $a=4.12$ \AA, $c=13.58$ \AA\ \cite{Zhao.K_etal.Nat-Commun2013} and the arsenic height $h_{\text{As}}=1.541$ \AA\ \cite{Hellmann.A_etal.Z.-Naturforsch.2007}.  First we tried the local density approximation (LDA) and generalized gradient approximation (GGA) exchange functionals \cite{Perdew.J_etal.Phys.-Rev.-Lett.1996,Perdew.J_etal.Phys.-Rev.-Lett.1997}, but they gave overlapping conduction and valence bands, inconsistent with the semiconducting electrical conductivity with a band gap of 0.23 eV \cite{Xiao.Z_etal.Journal-of-the-American-Chemical-Society2014}. This is the well-known underestimation of band gap in LDA and GGA, and therefore we next employed the so-called modified Becke-Johnson exchange potential (mBJ) implemented in Wien2k program \cite{Tran.F_etal.Phys.-Rev.-B2011} with a standard mixing factor for the exact-exchange term of 0.25 \cite{Perdew.J_etal.The-Journal-of-Chemical-Physics1996}. We thus obtained a semiconducting band structure with the valence-band maximum at the $\Gamma$ point and  the conduction-band minimum at the Z point, in agreement with the previous calculation using optimized atomic positions \cite{Shein.I_etal.Journal-of-Alloys-and-Compounds2014}.
A significant reduction of the resistivity at $T=0$ through K substitution to BaZn$_{2}$As$_{2}$ \cite{Zhao.K_etal.Nat-Commun2013}, from $1\times 10^{4}$ $\Omega$ cm ($x=0$) to $5\times 10^{-1}$ $\Omega$ cm ($x=0.1$), is caused by the downward shift of $E_{F}$. Unlike the iron pnictide superconductors, the conduction bands are composed mainly of Ba $5d$ and As $4d$ orbitals, and the valence bands are composed of the Zn $4p$ and As $4p$ orbitals. Weakly dispersive bands around -7.5 eV originate from the Zn $3d$ orbitals and have a DOS as high as 50 eV$^{-1}$ at the peak position. 
   
\begin{figure}[htbp]
\begin{center}
\includegraphics[width=8cm]{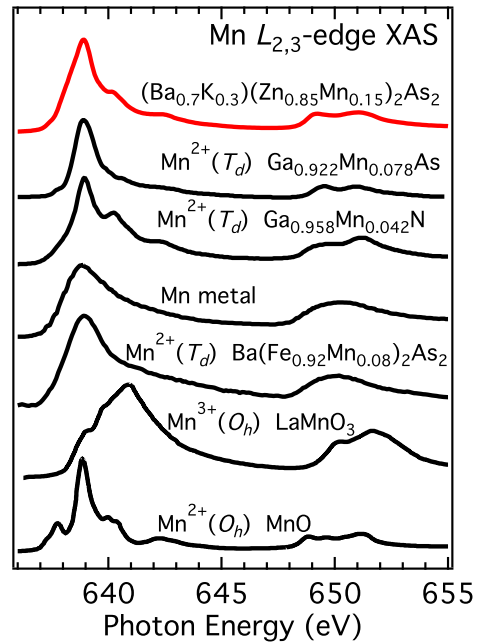}
\caption{(Color online) Mn $L_{2,3}$-edge XAS spectra of Ba$_{0.7}$K$_{0.3}$(Zn$_{0.85}$Mn$_{0.15}$)$_{2}$As$_{2}$ (Mn-BaZn$_{2}$As$_{2}$). The spectrum is compared with those of Ga$_{0.922}$Mn$_{0.078}$As \cite{Takeda.Y_etal.Phys.-Rev.-Lett.2008}, Ga$_{0.958}$Mn$_{0.042}$N \cite{Hwang.J_etal.Applied-Physics-Letters2007}, Mn metal \cite{Andrieu.S_etal.Phys.-Rev.-B1998}, Ba(Fe$_{0.92}$Mn$_{0.08}$)$_{2}$As$_{2}$ \cite{Suzuki.H_etal.Phys.-Rev.-B2013}, LaMnO$_{3}$, and MnO \cite{Burnus.T_etal.Phys.-Rev.-B2008}. The valence and the local symmetry of the Mn atom are indicated for each compound.}
\label{XAS}
\end{center}
\end{figure}

In order to clarify the electronic states of the doped Mn, we have performed XAS measurements in the photon energy regions around the $L_{2,3}$ edge of Mn. Figure \ref{XAS} shows the Mn $L_{2,3}$ absorption edges of Ba$_{0.7}$K$_{0.3}$(Zn$_{0.85}$Mn$_{0.15}$)$_{2}$As$_{2}$ compared with those of some reference systems. 
The line shapes of Ba$_{0.7}$K$_{0.3}$(Zn$_{0.85}$Mn$_{0.15}$)$_{2}$As$_{2}$ is intermediate between two DMS systems Ga$_{0.922}$Mn$_{0.078}$As \cite{Takeda.Y_etal.Phys.-Rev.-Lett.2008} (GaMnAs) and Ga$_{0.958}$Mn$_{0.042}$N \cite{Hwang.J_etal.Applied-Physics-Letters2007} (GaMnN), indicating that the Mn atoms take the valence of 2+ and that Mn $3d$ orbitals strongly hybridize with the surrounding As $4p$ orbitals as in GaMnAs and GaMnN. From the shoulder structures around $h\nu $ $=$ 640 and 643 eV, which are more pronounced than in GaMnAs and weaker than in GaMnN,  we see that the strength of hybridization is weaker than in GaMnAs but stronger than in GaMnN. The line shape has more localized nature than metallic compounds Mn metal \cite{Andrieu.S_etal.Phys.-Rev.-B1998} and Mn doped into BaFe$_{2}$As$_{2}$ \cite{Suzuki.H_etal.Phys.-Rev.-B2013}. However, it does not have clear multiplet structures seen in the spectra of LaMnO$_{3}$ and MnO \cite{Burnus.T_etal.Phys.-Rev.-B2008}, consistent with the semi-metallic conductivity in Mn-BaZn$_{2}$As$_{2}$. 

\begin{figure}[htbp]
\begin{center} 
\includegraphics[width=8cm]{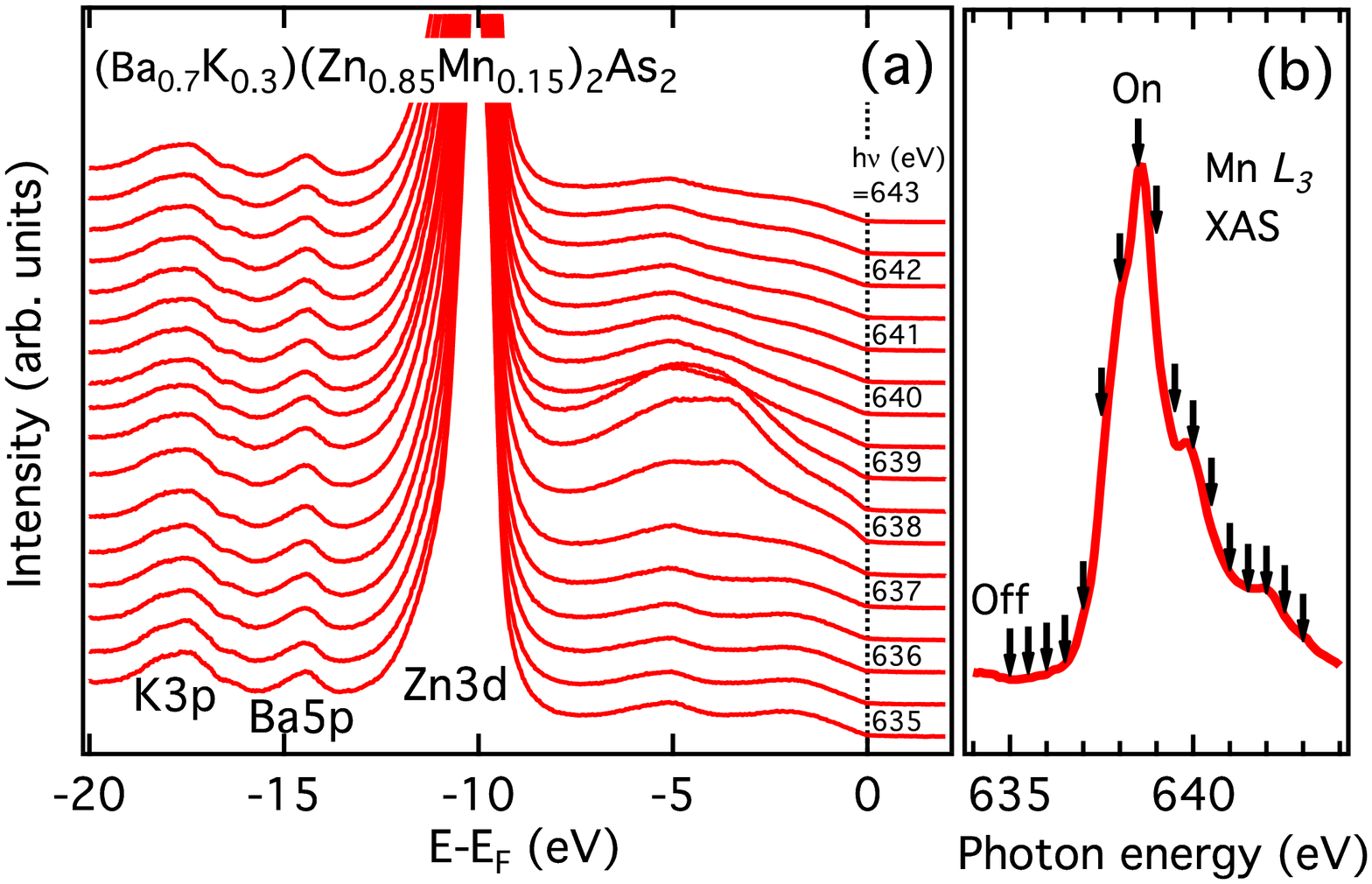}
\caption{(Color online) (a) Evolution of the valence-band photoemission spectra of Mn-BaZn$_{2}$As$_{2}$ with photon energy $h\nu=$ 635-643 eV. Excitation photon energies are shown by arrows on the XAS spectrum in panel (b).}
\label{Series}  
\end{center}
\end{figure}

In order to extract the local electronic structure of the doped Mn, we performed RPES experiments using photon energies around the Mn $L_{3}$ edge. In RPES, one makes use of the property that the cross-section of photoemission from an atomic orbital is enhanced by quantum-mechanical interference between direct photoemission of a $d$ electron, $3d^{n}+h\nu\rightarrow 3d^{n-1}+e^{-}$, and absorption followed by a Coster-Kr$\ddot{\text{o}}$nig transition $2p^{6}3d^{n}+h\nu\rightarrow 2p^{5}3d^{n+1}\rightarrow 2p^{6}3d^{n-1}+e^{-}$ \cite{Gelmukhanov.F_etal.Physics-Reports1999,Bruhwiler.P_etal.Rev.-Mod.-Phys.2002}. This effect is useful in extracting the $3d$ partial density of states (PDOS) of a transition element in solids. Following the observed x-ray absorption spectra, we measured the valence-band photoemission spectra with photon energies from the off-resonance to on-resonance regions. A series of photoemission spectra taken with a small photon energy interval enables us to clearly identify  the resonance enhancement of Mn $3d$-related photoemission features. 
  
Figure \ref{Series} (a) shows the valence-band spectra taken with photon energies in the Mn $L_{3}$ absorption region. Photon energies used are shown by arrows on the XAS spectrum in panel (b).
The high DOS of the Zn $3d$ states is clearly observed $\sim$ 10 eV below $E_{F}$. Note that, regardless of hole doping in Mn-BaZn$_{2}$As$_{2}$, the Zn $3d$ peak is located at $\sim$ -10 eV. (In DFT calculation, it is calculated to be $\sim$ -7.5 eV, see Fig. \ref{LDAband}. Similar discrepancy of the Zn $3d$ energy level between DFT and photoemission is also found in Zn-doped BaFe$_{2}$As$_{2}$ \cite{Berlijn.T_etal.Phys.-Rev.-Lett.2012,Ideta.S_etal.Phys.-Rev.-B2013}.) Photon-energy-independent peaks observed at -15 eV and -18 eV originate from the Ba $5p$ and K $3p$ orbitals, respectively. Most importantly, one can see the enhancement of spectral features in -8 $\sim$ 0 eV as the photon energy approaches on-resonance energy at 638.5 eV and the subsequent reduction of spectral weight at higher photon energies.
 
\begin{figure}[htbp]
\begin{center} 
\includegraphics[width=8cm]{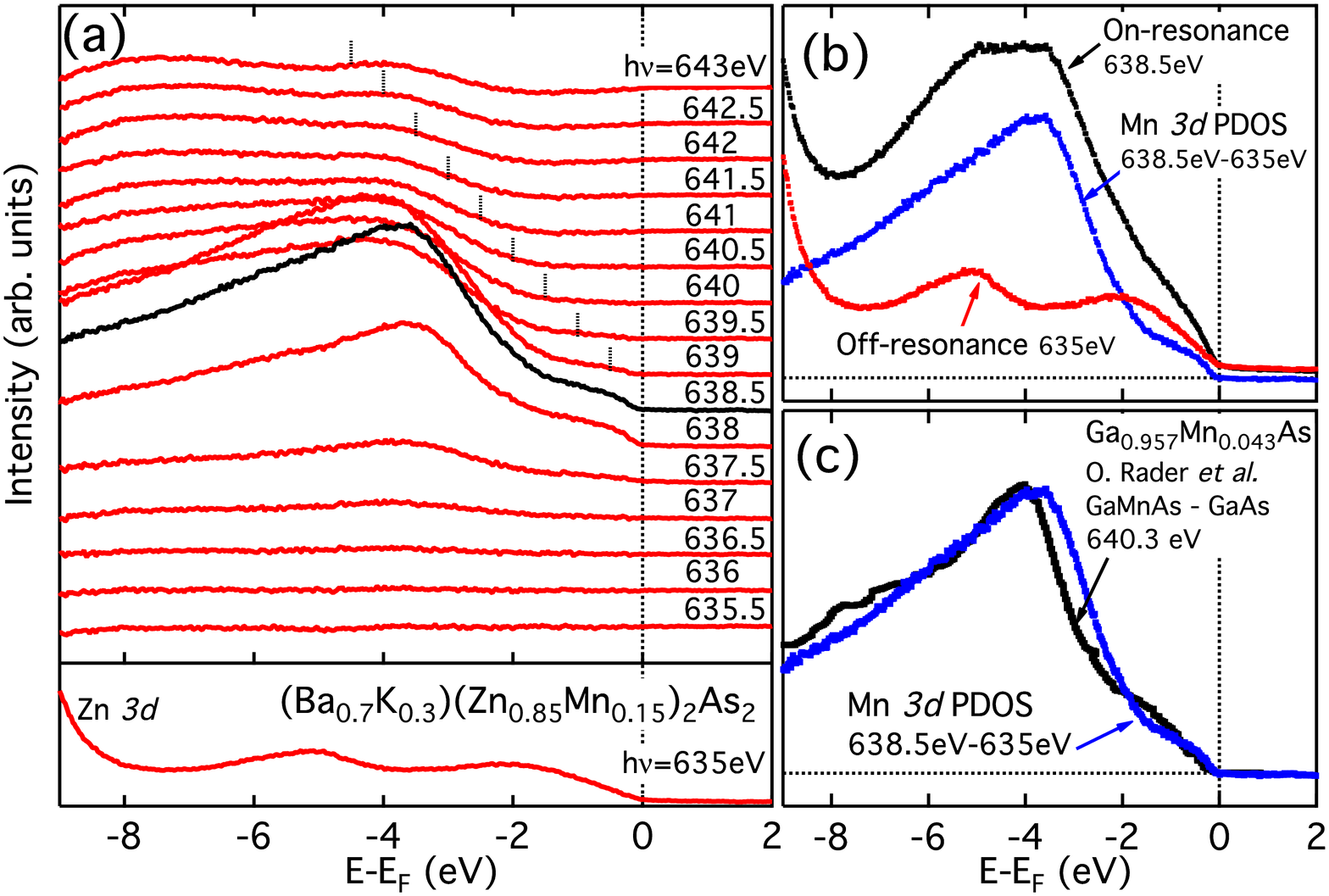}
\caption{(Color online) (a) Evolution of the valence-band photoemission difference spectra of Mn-BaZn$_{2}$As$_{2}$ for photon energies $h\nu=$ 635-643 eV. Off-resonance photoemission spectrum at the bottom of the panel ($h\nu$ = 635 eV) has been subtracted from the original spectra in order to highlight the resonant enhancement of the spectral weight. On-resonance spectrum is shown by a black curve. Vertical bars indicate a constant kinetic energy characteristic of Auger-electron emission if it existed. The absence of clear Auger peaks reflects the localized nature of the Mn $3d$ electrons. (b) Mn 3$d$ partial density of states  deduced by subtracting the off-resonance spectrum ($h\nu$ = 635 eV) from the on-resonance spectra ($h\nu$ = 638.5 eV). (f) Mn 3$d$ PDOS of Ga$_{0.957}$Mn$_{0.043}$As \cite{Rader.O_etal.Phys.-Rev.-B2004} compared with that of Mn-BaZn$_{2}$As$_{2}$.}
\label{RPES}  
\end{center}
\end{figure}
 
In order to highlight the resonance enhancement of Mn $3d$-derived spectral weight, we show on-off the difference spectra in Fig. \ref{RPES}. The off-resonance spectrum ($h\nu$ = 635 eV) at the bottom of Fig. \ref{RPES} (a) has been subtracted from each spectrum. 
The strongest enhancement around the Mn $L_{3}$ edge ($h\nu$ = 638.5 eV) is seen around -4 eV. Vertical bars indicate a constant kinetic energy characteristic of Auger emission. The absence of clear Auger peaks represents that the core hole created in the Mn $2p$ level is not efficiently screened before the Coster-Kr$\ddot{\text{o}}$nig decay due to the low Mn $3d$ PDOS around $E_{F}$ (below). From this result, we see that the Mn 3$d$ electrons are essentially localized and do not form band states with the As $4p$ valence band. The absence of Auger emission is similar to RPES spectra taken at Mn $L_{3}$ edges in Mn-doped BaFe$_{2}$As$_{2}$ \cite{Suzuki.H_etal.Phys.-Rev.-B2013}, which demonstrates the localized nature of Mn $3d$ electrons in the metallic FeAs layer, and is in contrast with the strong Auger feature observed at the Co $L_{2,3}$ edges in Ca(Fe$_{0.944}$Co$_{0.056}$)$_{2}$As$_{2}$ \cite{Levy.G_etal.Phys.-Rev.-Lett.2012}, which clearly signifies the metallic nature of Co 3$d$ electrons and the high Co 3$d$ PDOS at $E_{F}$. 

By subtracting the off-resonance spectra from the on-resonance spectra, we have deduced the PDOS of Mn $3d$ orbitals as shown in Fig. \ref{RPES} (b). The DOS is low at $E_{F}$, finite between -2 eV and $E_{F}$ and takes a maximum at -4 eV. Mn $3d$ spectral weight is widely distributed from $\sim$ -10 eV to $\sim$ -2 eV.  The deduced PDOS is compared with that of Ga$_{0.957}$Mn$_{0.043}$As \cite{Rader.O_etal.Phys.-Rev.-B2004} in Fig. \ref{RPES} (c). Except that the peak in Ga$_{0.957}$Mn$_{0.043}$As is about 0.4 eV deeper than that of Mn-BaZn$_{2}$As$_{2}$, the overall spectral shapes are quite similar, indicating that the electronic states of doped Mn are alike in these two DMS systems. This similarity originates from the same chemical valence 2+ of the Mn atoms and the tetrahedral coordination by the As $4p$ orbitals.

From the obtained energy levels, we gain insight into the location of hole carriers doped by K substitution in Mn-BaZn$_{2}$As$_{2}$. The Zn $3d$ orbital is located as deep as 10 eV below $E_{F}$ and thus cannot accommodate holes. Also, since the Mn $3d$ PDOS has a maximum at -4 eV and very little contribution at $E_{F}$, they cannot accept holes either. Therefore, holes are predominantly introduced into the valence band composed mainly of the As $4p$ states. Thus Mn atoms have the valence of 2+ as observed in the XAS spectrum in Fig. \ref{XAS}, and the local magnetic moments with $S=5/2$ are formed there in the presence of Hund's coupling between electrons of $d^{5}$ configuration. 

The formation of the local magnetic moments affects the hole mobility. Upon K doping into the parent compound ($x=y=0$), the resistivity at $T=0$ significantly decreases from $1\times 10^{4}$ $\Omega$ cm ($x=0$) to $5 \times 10^{-1}$ $\Omega$ cm ($x=0.1$) \cite{Zhao.K_etal.Nat-Commun2013}. On the other hand, it does not radically decrease in samples with $y=0.1$, from $4$ $\Omega$ cm ($x=0$) to $2$ $\Omega$ cm ($x=0.3$). This behavior indicates that, while hole carriers introduced into the As $4p$-derived valence band are originally itinerant, they are weakly bound to the local magnetic moments and lose mobility in the Mn-doped samples. 

The ferromagnetic correlation between the local $S=5/2$ spins of Mn$^{2+}$ ions is considered to be mediated by the holes weakly bound to them. In generating ferromagnetism, the positions of the impurity level and $E_{F}$ are key parameters. A combined approach of density functional theory and quantum Monte Carlo method \cite{Ohe.J_etal.Journal-of-the-Physical-Society-of-Japan2009} on the Haldane-Anderson model \cite{Anderson.P_etal.Phys.-Rev.1961,Haldane.F_etal.Phys.-Rev.-B1976} showed that, when $E_{F}$ is located between the bound impurity state and the valence-band maximum of GaMnAs, a long-range ferromagnetic correlation between the impurities develops as a result of the antiferromagnetic impurity-host coupling and that it is enhanced with decreasing temperature. Indeed, a recent resonance angle-resolved photoemission study on GaMnAs \cite{Kobayashi.M_etal.Phys.-Rev.-B2014} reveals the presence of a nondispersive impurity band in the vicinity of the valence-band maximum as a split-off Mn-impurity state. Given that the valence-band top of BaZn$_{2}$As$_{2}$ is higher than that of GaAs, Mn-BaZn$_{2}$As$_{2}$ seems to satisfy the above condition. However, in order to precisely determine the mechanism of ferromagnetism in Mn-BaZn$_{2}$As$_{2}$, it is essential to experimentally investigate the properties of the impurity state. Optical and angle-resolved photoemission spectroscopy studies on the new series of DMSs are desired in future studies.

In conclusion, by using XAS and RPES techniques, we have studied the electronic structure of Ba$_{1-x}$K$_{x}$(Zn$_{1-y}$Mn$_{y}$)$_{2}$As$_{2}$, in particular that related to the Mn $3d$ states. Mn $L_{2,3}$-edge XAS spectrum indicates that the doped Mn atoms have the valence 2+ and are strongly hybridized with the As $4p$ orbitals as in archetypal DMSs GaMnAs. 
The Mn $3d$ PDOS obtained by RPES shows a peak around $E_{B}=4$ eV and is relatively high between $E_{B}=0$-$2$ eV with little contribution at $E_{F}$. These electronic states below $E_{F}$ leads to the $d^{5}$ electron configuration of Mn atoms with the local magnetic moment of $S=5/2$.
From comparison between DFT band dispersions of the host semiconductor BaZn$_{2}$As$_{2}$ and the experimental Mn $3d$ PDOS of Mn-BaZn$_{2}$As$_{2}$, we conclude that doped holes go into the top of the As $4p$-derived valence band and are weakly bound to the Mn local spins. The ferromagnetic correlation between the local spins mediated by the hole carriers induces ferromagnetism in Ba$_{1-x}$K$_{x}$(Zn$_{1-y}$Mn$_{y}$)$_{2}$As$_{2}$.  

This work was supported by a Grant-in-Aid for Scientific Research from the JSPS (S22224005), Reimei Project from Japan Atomic Energy Agency, 
Friends of Todai Foundation in New York, and the US NSF PIRE (Partnership for International Research and Education: OISE-0968226). Work at IOPCAS was supported by NSF \& MOST of China through Research Projects. XAS experiment at NSRRC was
performed under the proposal No. 2012-2-086-4. Experiment at Photon Factory was approved by the Photon Factory Program Advisory Committee (Proposals No. 2011S2-003 and 2014G177). H.S. acknowledges financial support from Advanced Leading Graduate Course for Photon Science (ALPS). 

\bibliography{SuzukiDMS}

\end{document}